\documentclass[%
 aip,
 amsmath,amssymb,
 reprint,%
]{revtex4-1}

\usepackage{graphicx}
\usepackage{dcolumn}
\usepackage{bm}

\usepackage[utf8]{inputenc}
\usepackage[T1]{fontenc}
\usepackage{mathptmx}

\begin{document}


\title{Diffusion Phenomena in a Mixed Phase Space}

\author{Matheus S. Palmero$^{1,2,3}$, Gabriel I. D\'iaz$^{3}$, Peter V. E. McClintock$^2$ and Edson D.\ Leonel$^{1}$}

\affiliation{$^1$Departamento de F\'isica, UNESP - Univ Estadual Paulista, Av. 24A, 1515, Bela Vista, 13506-900, Rio Claro, SP, Brazil\\
$^2$ Department of Physics, Lancaster University, Lancaster, LA1 4YW, United Kingdom\\
$^3$ Instituto de F\'isica, IFUSP - Universidade de S\~ao Paulo, Rua do Mat\~ao, Tr.R 187, Cidade Universit\'aria, 05314-970, S\~ao Paulo, SP, Brazil\\
}

\pacs{05.45.-a, 05.45.Jc, 47.52.+j}

\begin{abstract}
We show that, in strongly chaotic dynamical systems, the average
particle velocity can be calculated analytically by consideration of
Brownian dynamics in phase space, the method of images and use of the classical diffusion equation. 
The method is demonstrated on the simplified Fermi-Ulam
accelerator model, which has a mixed phase space with chaotic seas,
invariant tori and Kolmogorov-Arnold-Moser (KAM) islands. The calculated
average velocities agree well with numerical simulations and with an
earlier empirical theory. The procedure can readily be extended to other
systems including time-dependent billiards.
\end{abstract}

\maketitle

\begin{quotation}

We present an approach to the analysis of complicated chaotic dynamical systems that are difficult to treat in other ways. It relies on the counter-intuitive application of a fundamental idea from classical continuum physics (probabilistic diffusion) to chaotic systems that are, of course, inherently deterministic. In particular, we consider diffusion and Brownian dynamics in the phase space of the chaotic system and show how the diffusion equation, applied in this unusual context, can provide an accurate description of the average velocity and its evolution. To demonstrate and validate the formalism, we take a well-known example from astrophysics - the Fermi-Ulam model.

\end{quotation}

\section{\label{sec1}Introduction}
The evolution of systems described by Hamiltonians with nonlinear
terms in their dynamical equations may exhibit either regularity or
chaos. The result is often a mixed phase space containing chaotic seas, invariant
tori and Kolmogorov-Arnold-Moser (KAM) islands \cite{lich}. Dynamical
systems with strong chaotic motion often exhibit diffusive behavior \cite{ott,chirikov}.
An intuitive example of this is to drop colored ink into water,
observing how the particles of ink move away from each other, spreading
out into the liquid. For a mixed phase space, however, an initial
condition e.g.\ around a KAM island may lead to very complicated
behavior. The stability structures influence directly the transport
properties of chaotic orbits \cite{venegeroles1}, often generating
so-called anomalous diffusion \cite{zasla,robnik}.

There are many scenarios where, rather than analysing the individual
behaviour of a single particle starting from a particular initial
condition, it is more interesting to consider the average properties of
the system, taking into account an ensemble of particles. Statistical
methods can then be used to describe the dynamical phenomena
\cite{venegeroles2,meiss,manchein}. Correspondingly, the properties and
construction of the phase space can lead to what are effectively
diffusion processes: as the dynamics evolves, there is diffusion of the
action, usually associated with the velocity of the particles, through
the phase space.

In this work we show that the classical diffusion equation
\cite{dif1,dif2,dif3,dif4,dif5} can be solved via a procedure well-known in electrostatics, namely the {\it method of images}, and used to describe the evolution of
the average velocity for a system characterised by a mixed phase space.
We will demonstrate the effectiveness and utility of this idea by
applying it to the well-known and widely-studied Fermi-Ulam model (FUM).

This paper is organised as follows. In Sec.\ \ref {sec2} we describe
the FUM, showing the nonlinear map associated with the dynamics and
introducing a picture of a diffusion process occurring within its
characteristic phase space. Section \ref{sec3} develops a theoretical
framework yielding analytical results for normal diffusion in a mixed
phase space. In Sec.\ \ref{sec4} we compare these analytical results
with numerical data. Conclusions are drawn in Sec.\ \ref{sec5}.

\section{\label{sec2}Model and phase space}
\begin{figure}[h!]
\centering
\includegraphics[width=8cm]{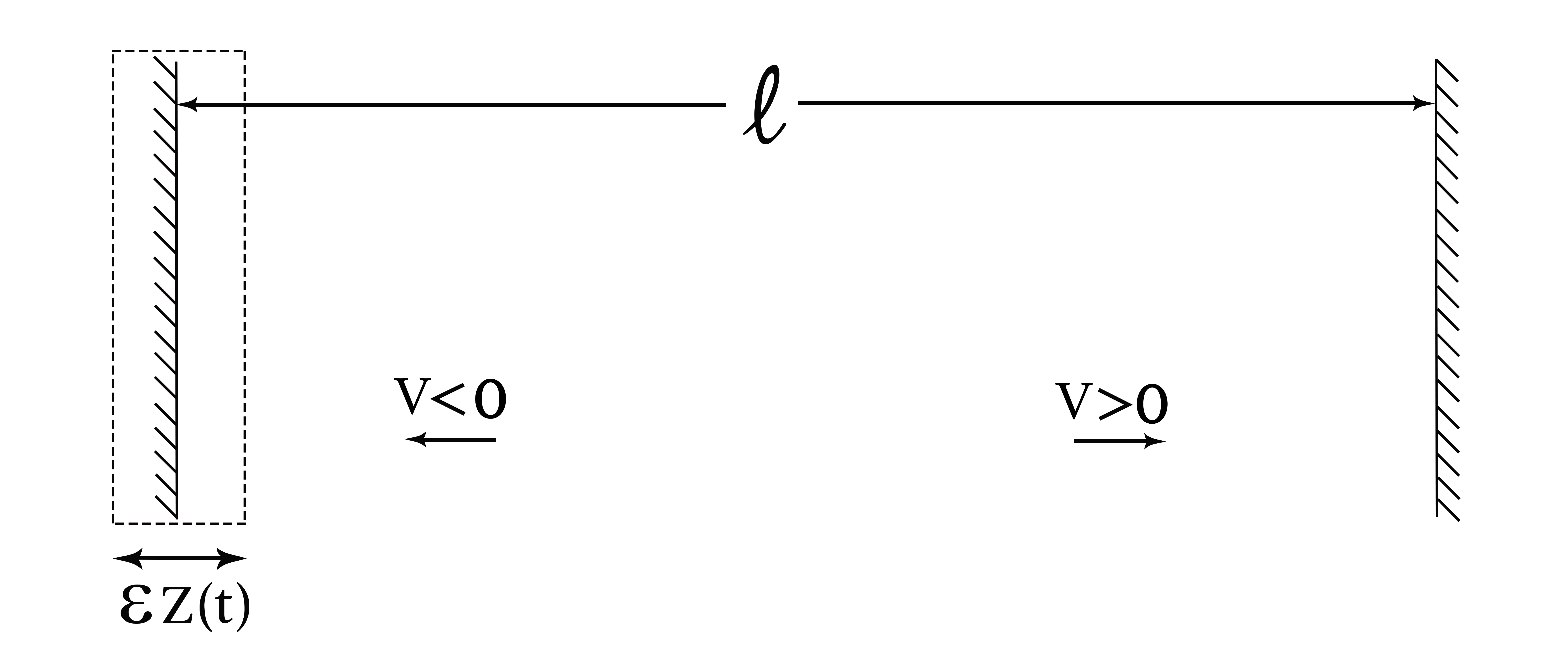}
\caption{Illustration of the Fermi-Ulam model. The geometrical
parameter $\ell$ is the distance between the two walls and the direction
of the vectors denotes the sign of the particle's velocity. Usually the
time-dependent function $Z(t)$ is chosen as $\cos(\omega t)$, with
$\omega$ the frequency of oscillation. }
\label{fig1}
\end{figure}

The FUM \cite{fum} is a version of the Fermi accelerator, which was
originally introduced by Enrico Fermi \cite{fermi} as a possible
explanation for the production of very high energy cosmic rays. Its
acceleration mechanism involves the repulsion of an electrically charged
particle by strong oscillatory magnetic fields, a process that is
analogous to a classical particle colliding with an oscillating physical
boundary. The model consists of a particle bouncing back and forth
between two rigid walls, one of which is fixed, whereas the other moves
periodically in time with a normalized amplitude $\varepsilon$, as shown
schematically in Fig.\ref{fig1}.

\begin{figure*}[t!]
\includegraphics[scale=0.22]{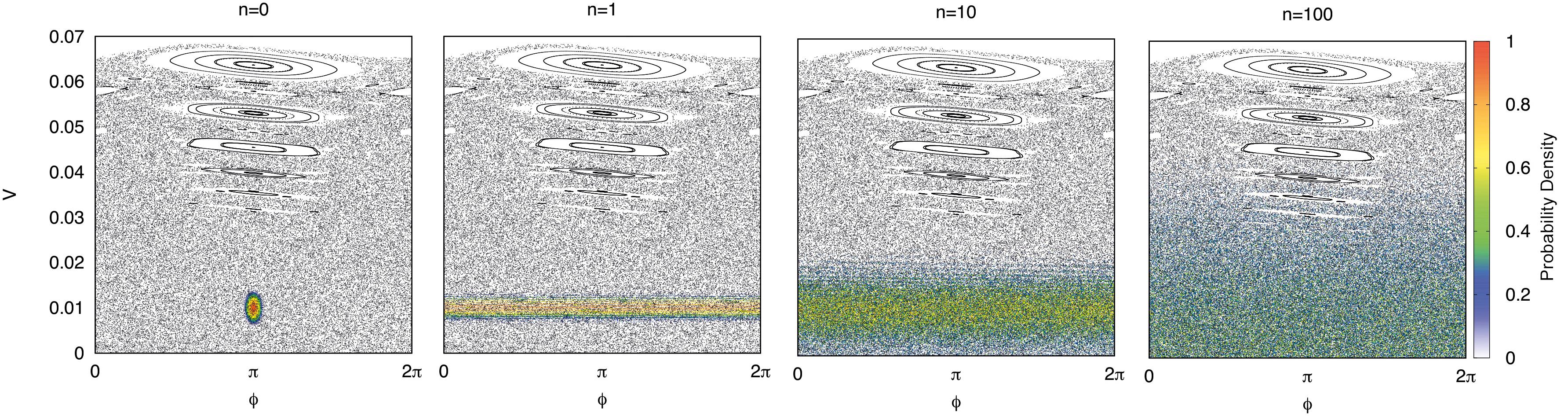}
\caption{(colour online). Phase space diffusion in the FUM, as sketched in Fig. \ref{fig1} and described by the mapping (\ref{eq1}). It is illustrated by the probability density in a chaotic region of the FUM's
phase space, for different numbers of
iterations $n$. The colour scale shows how likely it is to find an orbit
at that area of the phase space. The initial distribution, centered at
$\phi_0=\pi$, $V_0=0.01$ with a standard deviation
$\sigma_{\phi_0}=0.05$ and $\sigma_{V_0}=0.001$, was plotted overlaying
the phase space generated for the same parameter. As $n$ increases, the
distribution instantly spreads out uniformly along the $\phi$ axis and also
diffuses, albeit more slowly, towards smaller and larger $V$.}
\label{fig2}
\end{figure*}

The system is described by a two-dimensional, nonlinear,
area-preserving map $T(V_n,\phi_n)=(V_{n+1},\phi_{n+1})$. The velocity
of the particle is the action variable and the phase, related to the
time-dependent boundary, is the angle variable. Taking into account that
the absolute value of velocity changes at the moment of each collision,
the mapping for the simplified version \footnote{We approximate the oscillating wall as fixed. In reality, of course, when the particle suffers a collision, it exchanges momentum appropriate to a moving wall, but this simplified version is valid when the nonlinear parameter $\varepsilon$ is relatively small.} of the FUM is
\begin{equation}
T:\left\{\begin{array}{ll}
V_{n+1} =|V_n - 2\varepsilon\sin(\phi_{n+1})|~~\\
\phi_{n+1} = [\phi_n + \frac{2}{V_n}] \mod(2\pi)\\
\end{array}
\right..
\label{eq1}
\end{equation}\noindent
The term $\frac{2}{V_n}$ corresponds to the time between collisions and
$- 2\varepsilon\sin(\phi_{n+1})$ gives the gain or loss of
velocity/energy in each collision.

The phase space $V \times \phi$ for the FUM is composed of chaotic seas
and KAM islands, and is accordingly classified as a mixed phase space.
In addition, it is bounded by an invariant spanning curve which plays
the role of a boundary: trajectories of lower velocity will never visit
a region above this curve, no matter how many times the trajectory is
iterated.

The average velocity of an ensemble of particles inside the FUM grows
initially \cite{tiago}, and then flattens off towards a plateau. This
velocity growth and saturation can be interpreted as involving a
diffusion process, albeit diffusion not in the physical space of the
FUM, but rather in its phase space. Figure \ref{fig2} shows how this
phase-space-diffusion behaves for different numbers of iterations $n$.
At $n=0$ we have the initial Gaussian-shaped distribution centered at
$[V_0=0.01,\phi_0=\pi]$; then, one iteration latter, the distribution
seems to have become spread out uniformly along the phase axis, a fact
that will be used later in the analytic approach. However, diffusion is
also starting on the action axis. After $10$ and then $100$ iterations
of the mapping Eq.\ (\ref{eq1}), the action/velocity is still continuing
its diffusion through phase space. For all panels of Fig.\ \ref{fig2},
$\varepsilon=0.001$.

It is important to bear in mind that the phase space diffusion is
limited down by null velocity and up by the first invariant spanning
curve. Its position is approximated by $V_{\rm f}\approx2\sqrt{\varepsilon}$.
The localization of such a curve can be
obtained by using a connection with the standard mapping
\cite{lich,vfisc}, which is written as
\begin{equation}
T:\left\{\begin{array}{ll}
I_{n+1} =I_n +K\sin(\theta_n),~~\\
\theta_{n+1} = [\theta_n + I_n], \mod(2\pi)\\
\end{array}
\right.
\label{eq2}
\end{equation}
where the parameter $K$ controls the intensity of the nonlinearity of
the mapping. There are two transitions in the standard mapping: (i)
integrability when $K=0$ to non-integrability for any $K\ne0$; and (ii)
a transition from local chaos when $K<K_c$ to global chaos for $K>K_c$.
The parameter $K_c=0.9716\ldots$ identifies the critical value of
control parameter where all of the invariant spanning curves are
destroyed, letting the dynamics diffuse unbounded in the $I$ direction.
This is exactly the transition we want to use in connection with the
FUM as an attempt to describe the localization of the first invariant
spanning curve. Above the curve in the FUM, one observes local chaos,
an infinity of other invariant spanning curves and eventually periodic
orbits. Below the first invariant spanning curve only chaos, periodic and quasi
periodic dynamics coexist, each one of them being visited as determined by the initial conditions.
The procedure to obtain $V_{\rm f}$ consists of describing the position of
the first invariant spanning curve in the FUM through a local description of
the standard mapping. Then a Taylor expansion (see Ref. \cite{vfisc}
for more details in a family of area preserving mappings) is made in
the first equation of mapping (\ref{eq1}) by using the fact that the invariant
spanning curve is written as $V_n=V_{\rm f}+\Delta V_n$ where $\Delta
V_n\ll V_{\rm f}$ is a small perturbation of the typical value
$V_{\rm f}$. A first order approximation leads to the expression
$V_{\rm f}\approx2\sqrt{\varepsilon}$.

\section{\label{sec3}Analytical procedure}
Essentially, the action variable $V$ is undergoing a diffusion process
within the bounded space $V \in [0,V_{\rm f}]$. This can be described by the
diffusion equation with no flux through its boundaries $\frac{\partial
\rho(0,t)}{\partial V}=\frac{\partial \rho(V_{\rm f},t)}{\partial V}=0,
\forall t>0$. Thus, the problem may be reduced to that of solving the
diffusion equation to obtain the probability density function
$\rho(V,t)$; once this has been integrated along the bounded space
$\langle V\rangle= \int_{0}^{V_{\rm f}} V\rho(V,t) dV$, it yields a
theoretical prediction for the average velocity of $\langle V\rangle$ of
particles inside the FUM.

The solution of the diffusion equation with no flux through the
boundaries can be obtained analytically by the method of images, as in
electrostatics \cite{crank}. Basically the idea is to treat the initial
Gaussian distribution as a point charge and the boundaries as conducting
planes. The solution will then be an infinite sum of Gaussian functions
centered at $V_0$, due to the infinity of images of the initial profile.

First, we consider a normal diffusion process in one dimension, with no
boundaries and with the initial condition $\rho(V,0)=\delta(V-V_0)$,
where $V_0$ is the initial velocity of the particles. The fundamental
solution of the diffusion equation is given by
\begin{equation}
\rho(V,t)=\frac{1}{\sqrt{4\pi Dt}}e^{\frac{-(V-V_0)^2}{4Dt}}~.
\label{eq3}
\end{equation}\noindent
It is a Gaussian function (normal distribution). Solutions of this type
are well-known and widely applicable in science
\cite{gauss1,gauss2,gauss3}, especially, in statistical physics. Normal
distributions are characterized by their mean value $\mu\equiv\langle
V\rangle$ and variance $\sigma^2 \equiv\langle V^2\rangle - \langle
V\rangle^2$. Likewise, the diffusion coefficient can be written as a
function of the time derivative of the variance
$D=\frac{1}{2}\frac{d\sigma^2}{dt}~~\rightarrow~~\sigma^2=2Dt$. The
solution can then be rewritten in terms of $\mu$ and $\sigma^2$ as
\begin{equation}
\rho(V;\mu,\sigma^2)=\frac{1}{\sqrt{2\pi
\sigma^2}}e^{\frac{-(V-\mu)^2}{2\sigma^2}}~.
\label{eq4}
\end{equation}\noindent
Knowing the fundamental solution, and applying the principle of
superposition as in the method of images, we may assume that a sum of
Gaussian functions is still a solution to the problem. Hence the
solution when $V \in [0,V_{\rm f}]$ with $\frac{\partial \rho(0,t)}{\partial
V}=\frac{\partial \rho(V_{\rm f},t)}{\partial V}=0$ is given by
\begin{small}
\begin{multline}
\rho(V;\mu,\sigma^2)=\frac{1}{\sqrt{2\pi
\sigma^2}}\sum_{m=-\infty}^{\infty}\left[\right.
\exp\left(\frac{-(V-2mV_{\rm f}-\mu)^2}{2 \sigma^2}\right)\\
+\exp\left(\frac{-(V-2mV_{\rm f}+\mu)^2}{2\sigma^2}\right)\left.\right].
\label{eq5}
\end{multline}
\end{small}\noindent
As it stands, however, this solution is not normalized for the space interval $V \in [0,V_{\rm f}]$. To effect normalization, it is necessary that $A \int_{0}^{V_{\rm f}} \rho(V;\mu,\sigma^2) dV~ = 1$, with $A$ equal to a normalization constant. Considering the error function property $\text{erf}(-x)=-\text{erf}(x)$, the normalized solution is given by
\begin{small}
\begin{multline}
\rho(V;\mu,\sigma^2)=\frac{1}{2A\sqrt{2\pi
\sigma^2}}\sum_{m=-\infty}^{\infty}\left[\right.
\exp\left(\frac{-(V-2mV_{\rm f}-\mu)^2}{2 \sigma^2}\right)\\
+\exp\left(\frac{-(V-2mV_{\rm f}+\mu)^2}{2\sigma^2}\right)\left.\right]~,
\label{eq6}
\end{multline}
\end{small}with \begin{small}
$A=\sum_{m=-\infty}^{\infty}\text{erf}\left(\frac{\mu-2mV_{\rm f}}{\sqrt{2\sigma^2
}}\right)-\text{erf}\left(\frac{\mu-V_{\rm f}-2mV_{\rm f}}{\sqrt{2\sigma^2}}\right)-\text{
erf}\left(\frac{\mu+2mV_{\rm f}}{\sqrt{2\sigma^2}}\right)+\text{erf}\left(\frac{
\mu+V_{\rm f}+2mV_{\rm f}}{\sqrt{2\sigma^2}}\right)$\end{small}.

\begin{figure}[h!]
\centering
\includegraphics[width=8.5cm]{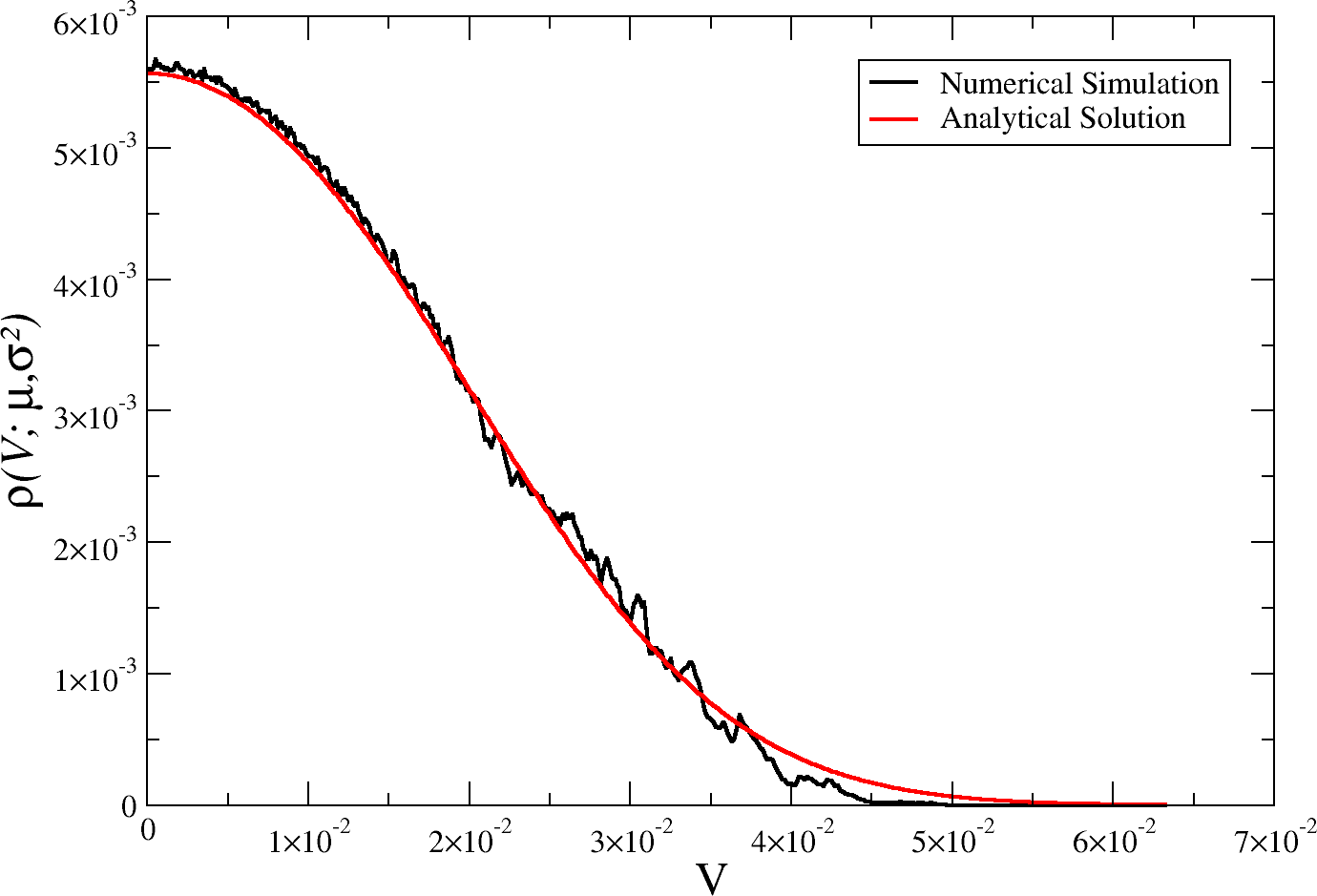}
\caption{Comparison between the analytical solution and the experimental/numerical probability distribution of the diffusion process depicted in Fig.\ \ref{fig2}. This is the behaviour after $100$ iterations.}
\label{fig3}
\end{figure}

Fig.\ \ref{fig3} shows how the analytical solution for the probability density given by Eq.\ (\ref{eq6}) fits the numerical simulation data for the FUM. The initial conditions for the analytic curve are the same as those used in constructing Fig.\ \ref{fig2}, with $V_0=0.01$, $\sigma_{V_0}=0.001$ and $\varepsilon=0.001$. This comparison \cite{robnik2} provides a convincing verification of the analytical solution. The good fit indicates that the solution is suitable when considering an initial profile in a chaotic region and neglecting the anomalous diffusion phenomena around KAM islands.

Having obtained this solution, we need to calculate the average as $\langle V\rangle= \int_{0}^{V_{\rm f}} x\rho(V;\mu,\sigma^2) dV$ in order to be able to predict analytically the average behaviour of the velocity. Because of the lack of symmetry, this calculation is non-trivial but, integrating between the upper and lower limits using the Jacobi Theta function representation \cite{jacobi}, we find that the solution can be written as
\begin{multline}
\langle V\rangle=\frac{1}{2A\sqrt{\pi}}\sum_{m=-\infty}^{\infty}\sqrt{2\sigma^2}(\Delta^{(1)}\text{exp}+\Delta^{(2)}\text{exp})\\
+ \sqrt{\pi}\left[(2mV_{\rm f}-\mu)\Delta^{(1)}\text{erf}+(2mV_{\rm f}+\mu)\Delta^{(2)}\text{erf}\right]~,
\label{eq7}
\end{multline}where

\begin{footnotesize}
\begin{eqnarray*}
\Delta^{(1)}\text{exp} &=& \exp\left(-\left(\frac{\mu-2mV_{\rm f}}{\sqrt{2\sigma^2}}\right)^2\right)-\exp\left(-\left(\frac{\mu-2mV_{\rm f}+V_{\rm f}}{\sqrt{2\sigma^2}}\right)^2\right),\\
\Delta^{(2)}\text{exp} &=& \exp\left(-\left(\frac{\mu+2mV_{\rm f}}{\sqrt{2\sigma^2}}\right)^2\right)-\exp\left(-\left(\frac{\mu+2mV_{\rm f}-V_{\rm f}}{\sqrt{2\sigma^2}}\right)^2\right),\\
\Delta^{(1)}\text{erf} &=& \text{erf}\left(\frac{\mu-2mV_{\rm f}-V_{\rm f}}{\sqrt{2\sigma^2}}\right)-\text{erf}\left(\frac{\mu-2mV_{\rm f}}{\sqrt{2\sigma^2}}\right),\\
\Delta^{(2)}\text{erf} &=& \text{erf}\left(\frac{\mu+2mV_{\rm f}}{\sqrt{2\sigma^2}}\right)-\text{erf}\left(\frac{\mu+2mV_{\rm f}+V_{\rm f}}{\sqrt{2\sigma^2}}\right).
\end{eqnarray*}
\end{footnotesize}\noindent
Defining an important auxiliary variable $z=\frac{\mu}{\sqrt{2\sigma^2}}$ and a new parameter $\tilde{v}=\frac{V_{\rm f}}{\sqrt{2\sigma^2}}$ making the necessary re-arrangements, the average velocity within the FUM is given analytically by
\begin{multline}
\langle V\rangle=\frac{\mu}{2A}\sum_{m=-\infty}^{\infty}\frac{1}{z\sqrt{\pi}}(\Delta^{(1)}\text{exp}+\Delta^{(2)}\text{exp}) \\
+ \frac{1}{\mu}\left[\left(\frac{2mV_{\rm f}}{\sqrt{2\sigma^2}}-\mu\right)\Delta^{(1)}\text{erf}+\left(\frac{2mV_{\rm f}}{\sqrt{2\sigma^2}}+\mu\right)\Delta^{(2)}\text{erf}\right]~,
\label{eq8}
\end{multline}with \begin{small}$A=\sum_{m=-\infty}^{\infty}\text{erf}\left(z-\frac{2mV_{\rm f}}{\sqrt{2\sigma^2}}\right)-\text{erf}\left(z-\tilde{v}-\frac{2mV_{\rm f}}{\sqrt{2\sigma^2}}\right)-\text{erf}\left(z+\frac{2mV_f}{\sqrt{2\sigma^2}}\right)+\text{erf}\left(z+\tilde{v}+\frac{2mV_{\rm f}}{\sqrt{2\sigma^2}}\right)$\end{small} and

\begin{small}
\begin{eqnarray*}
\Delta^{(1)}\text{exp} &=& e^{-\left(z-\frac{2mV_{\rm f}}{\sqrt{2\sigma^2}}\right)^2}-e^{-\left(z-\frac{2mV_{\rm f}}{\sqrt{2\sigma^2}}+\tilde{v}\right)^2},\\
\Delta^{(2)}\text{exp} &=& e^{-\left(z+\frac{2mV_{\rm f}}{\sqrt{2\sigma^2}}\right)^2}-e^{-\left(z+\frac{2mV_{\rm f}}{\sqrt{2\sigma^2}}-\tilde{v}\right)^2},\\
\Delta^{(1)}\text{erf} &=& \text{erf}\left(z-\frac{2mV_f}{\sqrt{2\sigma^2}}+\tilde{v}\right)-\text{erf}\left(z-\frac{2mV_{\rm f}}{\sqrt{2\sigma^2}}\right),\\
\Delta^{(2)}\text{erf} &=& \text{erf}\left(z+\frac{2mV_{\rm f}}{\sqrt{2\sigma^2}}\right)-\text{erf}\left(z+\frac{2mV_{\rm f}}{\sqrt{2\sigma^2}}+\tilde{v}\right).
\end{eqnarray*}
\end{small}\noindent

Furthermore, the mean $\mu$ and variance $\sigma^2$ are calculated, by construction, over the {\it point charge}, which is characterised by an unbounded diffusion process. According to our initial mapping, Eq.\ (\ref{eq1}), the point charge mapping is given by
$V_{n+1}=V_n-2\varepsilon\sin(\phi)$, where $\phi$ is an uniform random variable, as observed in Fig.\ \ref{fig2}, it is then
possible to write the mean and variance for the point charge as

\begin{equation*}
 \begin{split}
 \mu_{n+1}&=\langle V_{n+1}\rangle=\langle V_n\rangle \Rightarrow \mu=\mu_0=V_0\\
 \sigma^2_{n+1}&=\langle V^2_{n+1}\rangle - \langle V_{n+1}\rangle^2 = \langle V^2_n\rangle +2\varepsilon^2 - \langle V_n\rangle^2 \\
& \Rightarrow \sigma^2_{n+1}=\sigma^2_n + 2\varepsilon^2~.
 \end{split}
\end{equation*}\noindent
Following the theory of difference equations \cite{aprox}, assuming a large number of iterations and small values of $\varepsilon$, it is then possible to write $\sigma$ as a function of $n$
\begin{equation*}
 \sigma^2_{n+1}-\sigma^2_n=\frac{d \sigma^2}{dn}~\Rightarrow~ \sigma^2(n)=\sigma^2_0 + 2\varepsilon^2 n~.
\end{equation*}\noindent
This result is important because it carries the information that the variance is a function of the number of iterations $\sigma^2=\sigma^2(n)$, connecting the solution of the diffusion equation to the discrete mapping of the FUM. Moreover, the initial variance $\sigma_0$ is zero if the initial profile is considered a perfect Dirac delta function. This also tells us the diffusion is normal, since $\sigma\propto \sqrt n$. In addition, it is also possible to calculate the diffusion coefficient, which is a constant and quite intuitive with our suppositions for this case, so that $D=\epsilon^2$. Then $z$ is also a function of the number of iterations $n$ such that

\begin{figure}[b!]
\includegraphics[scale=0.45]{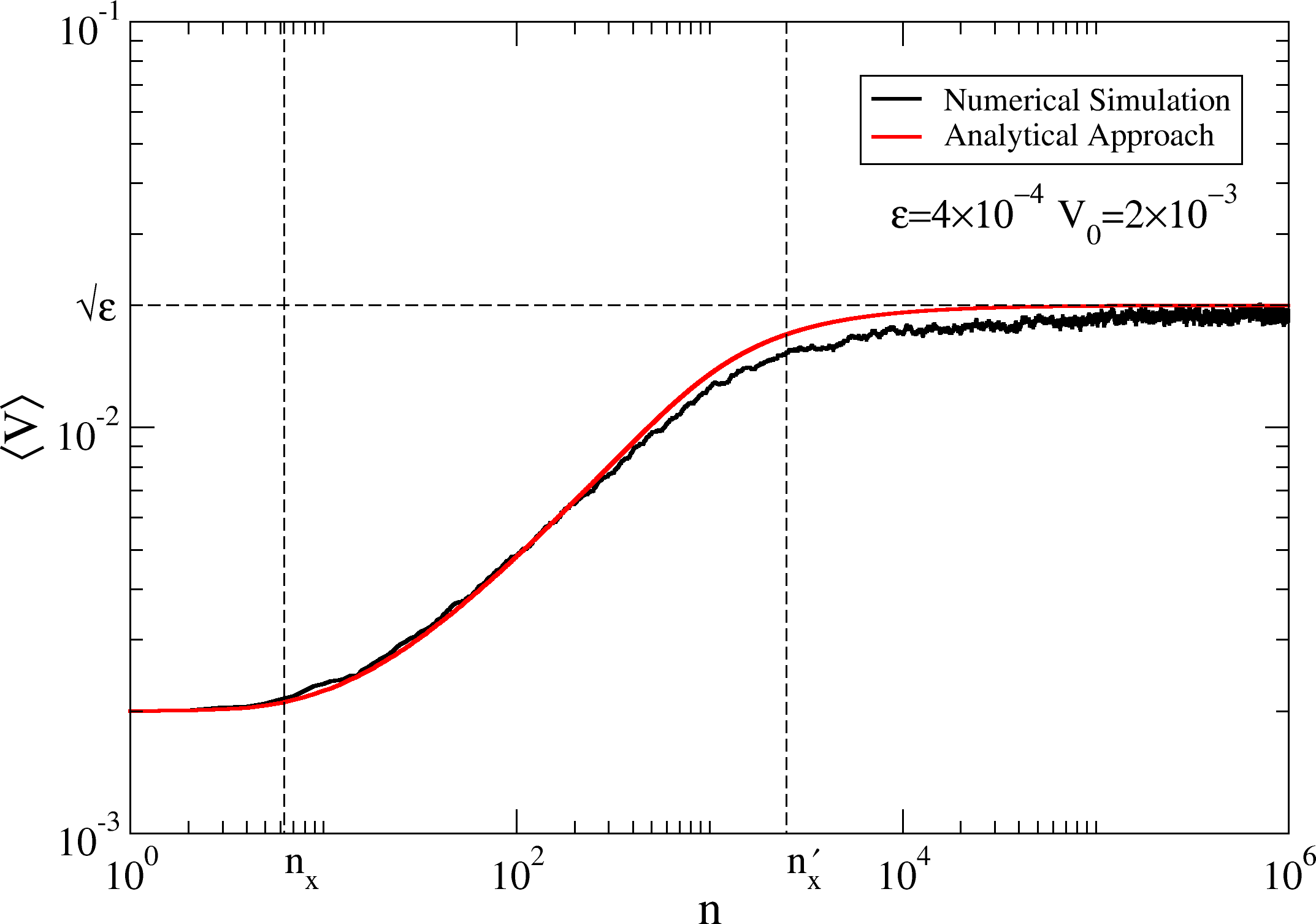}
\caption{(colour online). The average velocity $\langle V \rangle$ of particles in the FUM showing its evolution with the number of iterations $n$. With parameter $\varepsilon=4\times10^{-4}$ an ensemble of $10^3$ particles, each with $V_0=2\times10^{-3}$ was iterated until there had been $10^6$ collisions. The numerical simulations (rough black line) are compared
with the analytic theory (smooth red line). Note the saturation of $\langle V \rangle$ towards $\sqrt{\varepsilon}$ that occurs at large $n$ in both theory and simulation.}
\label{fig4}
\end{figure}

\begin{equation}
z=\frac{\mu}{\sqrt{2}\sigma(n)} ~ \Rightarrow ~ z(n)=\frac{V_0}{2\varepsilon\sqrt{n}}~.
\label{eq9}
\end{equation}

\noindent Substituting Eq.\ (\ref{eq9}) into Eq.\ (\ref{eq8}), we can calculate how the average velocity behaves as a function of the number of iterations for the dynamics of the FUM. But of course we now need to check whether, or not, this theory really describes the actual behaviour of the average velocity.

\section{\label{sec4}Analytical $\times$ Numerical results}

Fig.\ \ref{fig4} compares the numerical simulation data with the analytic predictions of Eqs.\ (\ref{eq8}),(\ref{eq9}). The expression for $\langle V\rangle$, given by Eq.\ (\ref{eq8}), represents a continuous competition between the exponential and error functions, so it is interesting to study their arguments. Based on a graphical analysis, we conclude that there are two changes of behaviour: at $z=1$; and at $\tilde{v}=1$. First, taking $z=1$
\begin{equation*}
z=1 ~ \Rightarrow ~ \frac{V_0}{2\varepsilon\sqrt{n}}=1 ~\Rightarrow ~ n=\left(\frac{V_0}{2\varepsilon}\right)^2 ~\Rightarrow~ n=\frac{V_0^2}{4\varepsilon^2}~,
\end{equation*}but in this case, $n=n_x$ marking the first crossover \footnote{The crossover number is defined as the iteration number that marks some dynamical change, for example the change from growth to saturation.}. Thus
\begin{equation}
n_x=\frac{V_0^2}{4\varepsilon^2}~.
\label{eq10}
\end{equation}\noindent
Secondly, taking $\tilde{v}=1$
\begin{equation*}
\tilde{v}=1 ~ \Rightarrow ~ \frac{V_{\rm f}}{2\varepsilon\sqrt{n}}=1 ~\Rightarrow ~ n\approx\left(\frac{2\sqrt{\varepsilon}}{2\varepsilon}\right)^2 ~\Rightarrow ~ n\approx\frac{1}{\varepsilon}~,
\end{equation*}but now, $n=n'_x$ marking the second crossover. Thus
\begin{equation}
n'_x\approx\frac{1}{\varepsilon}~.
\label{eq11}
\end{equation}\noindent
Another important result is the limit
\begin{equation}
\lim_{\sigma \rightarrow \infty}\langle V\rangle =\lim_{n \rightarrow \infty}\langle V\rangle = \frac{V_{\rm f}}{2} \approx \sqrt{\varepsilon}~,
\label{eq12}
\end{equation}which provides the saturation value $V_{sat}$.
\noindent Then, Fig.\ \ref{fig4} shows the average velocity for an ensemble of $10^3$ particles, all with initial velocity $V_0=2\times10^{-3}$, taken within the interval $\phi_0\in[0,2\pi]$. The analytical predictions for the first crossover $n_x$, Eq.\ (\ref{eq10}), the second crossover $n'_x$, Eq.\ (\ref{eq11}), and the saturation plateau when $n\rightarrow\infty$, Eq.\ (\ref{eq12}), are shown by the dashed lines.

The analytical approach, yielding Eq.\ (\ref{eq8}), clearly agrees well with the numerical simulation data. The correspondence might have been even closer were if not for the fact that the diffusion is not ideal for higher values of $V$, due to the configuration of the phase space. This also explains the fluctuation for $n>n'_x$. The diffusion around stability structures like KAM islands leads to the very complicated behaviour known as anomalous diffusion. However, the associated stickiness of the dynamics near the islands, though real, is a relatively minor effect given the size of the whole phase space: Harsoula {\it et al} \cite{harsoula} conclude that, for a long enough interval, averaging over the ensemble smooths the observables so that the stickiness can largely be neglected.

\begin{figure}[h!]
\includegraphics[scale=0.18]{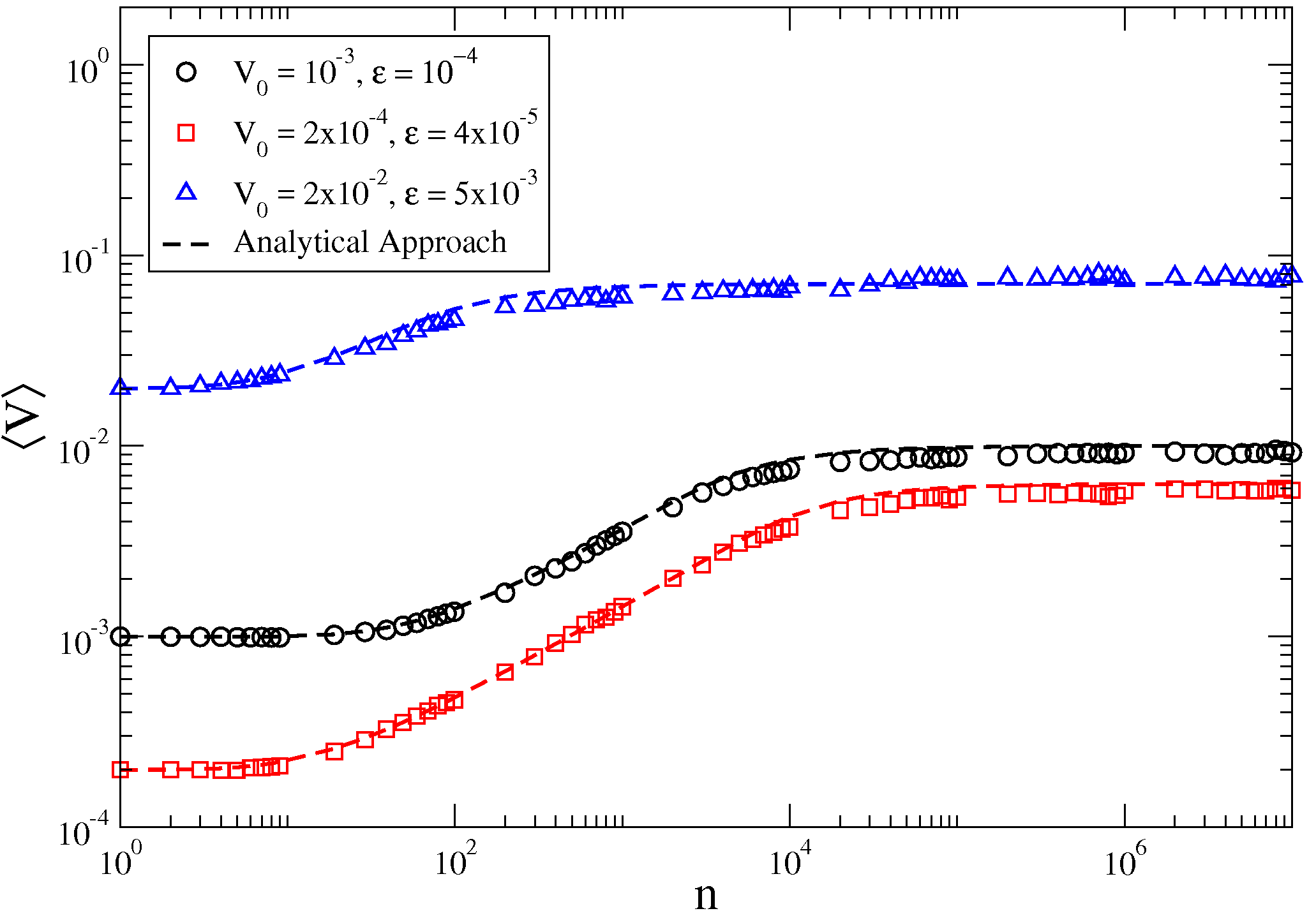}
\caption{(colour online). The average velocity $\langle V \rangle$ of particles in the FUM showing how it evolves with the number of iterations $n$, under different conditions. The analytic theory (dashed lines) is compared with numerical simulations (data points) for three different initial velocities $V_0$ and values of the control parameter $\varepsilon$, as listed in the inset. In each case, the simulations involved an ensemble of $10^4$ particles iterated up to $10^7$ collisions.}
\label{fig5}
\end{figure}

Fig.\ \ref{fig5} shows compares numerical data with the corresponding analytical predictions for three different initial velocities $V_0$ and values of the control parameter $\varepsilon$. It is important to remember that the position of the upper boundary in the phase space, which is the first invariant spanning curve, is approximated by $V_{\rm f}\approx2\sqrt{\varepsilon}$. Then, for each value of the parameter $\varepsilon$, a different bounded phase space is considered. Again, it is evident that the analytic curves provide an excellent fit to the numerical data, even for relatively large values of $\varepsilon$.

We emphasize that Eqs.\ (\ref{eq10}, \ref{eq11}, \ref{eq12}) represent the first analytic predictions to be made for the Fermi-Ulam model. They agree well with what was proposed on purely empirical grounds \cite{prl2004} more than a decade ago. Three hypotheses were then proposed, based on a scaling analysis: (i) $n_x \propto \frac{V_0^2}{\varepsilon^2}$ which agrees perfectly with Eq.\ (\ref{eq10}), and we now also obtain the proportionality constant $\frac{1}{4}$; in addition (ii) $n'_x \propto \frac{1}{\varepsilon}$ which agrees with Eq.\ (\ref{eq11}); and finally (iii) $V_{sat} \propto \varepsilon^{\alpha}$, with $\alpha\approx\frac{1}{2}$, which agrees with Eq.\ (\ref{eq12}).

\section{\label{sec5}Conclusion}
We conclude that a combination of the theory of diffusive processes with dynamical systems theory, plus the method of images from electrostatics, provides a powerful method for treating systems described by nonlinear mappings. The method can be expected to work for mixed phase spaces that are delimited by boundaries through which there are no fluxes. Application to the Fermi-Ulam model, taken as an example, has yielded some interesting features and excellent agreement both with numerical simulations and with earlier empirically-based theoretical considerations. Extension of the procedure discussed here to time-dependent billiards \cite{gelfreich} is an interesting possibility for future work.

\begin{acknowledgments}
We gratefully acknowledge valuable discussions with Professor Roberto Lagos. M. S. Palmero was supported by Funda\c c\~ao de Amparo \`a Pesquisa do Estado de S\~ao Paulo FAPESP, from Brazil, processes number 2014/27260-5 and 2016/15713-0. G. D. Iturry thanks to Brazilian agency CNPq. The research was supported by the Engineering and Physical Sciences Research Council (United Kingdom) under grants GR/R03631 and EP/M015831/1. E. D. Leonel acknowledges support from CNPq (303707/2015-1) and FAPESP (2017/14414-2). 
\end{acknowledgments}


\begin{thebibliography}{99}
\bibitem{simpliver} Suppose that the oscillating wall is fixed, however when the particle suffers a collision, it exchanges momentum as if the wall were moving. This simplified version is indeed valid when the nonlinear parameter $\varepsilon$ is relatively small.
\end{thebibliography}

\end{document}